\documentclass[prl,showpacs,aps,twocolumn,superscriptaddress,floatfix]{revtex4}

\usepackage{color}

\usepackage{graphicx}
\usepackage{subfigure}
\usepackage{bbold}
\usepackage{verbatim}
\usepackage{float}
\usepackage{enumerate}
\usepackage{amsfonts,amsmath}
\usepackage{multirow}
\usepackage[colorlinks,bookmarks=false,citecolor=blue,linkcolor=red,urlcolor=blue]{hyperref}

\begin{document}

\title{Putative spin--nematic phase in BaCdVO(PO$_{4}$)$_{2}$}

\author{M.~Skoulatos}
\affiliation{Heinz Maier-Leibnitz Zentrum (MLZ) and Physics Department E21, Technische Universit\"at M\"unchen, D-85748 Garching, Germany}
\affiliation{Laboratory for Neutron Scattering and Imaging, Paul Scherrer Institute, CH--5232 Villigen, Switzerland}

\author{F.~Rucker}
\affiliation{Physics Department E51, Technische Universit\"at M\"unchen, D-85748 Garching, Germany}

\author{G.J.~Nilsen}
\affiliation{Institut Laue-Langevin, 6 rue Jules Horowitz, 38042 Grenoble, France}
\affiliation{ISIS Neutron and Muon Facility, Rutherford Appleton Laboratory, Didcot OX11 0QX, United Kingdom}

\author{A.~Bertin}
\affiliation{Institut f\"ur Festk\"orperphysik, TU Dresden, D-01062, Dresden, Germany}

\author{E.~Pomjakushina}
\affiliation{Laboratory for Multiscale Materials Experiments, Paul Scherrer Institute, CH--5232 Villigen, Switzerland}

\author{J.~Ollivier}
\affiliation{Institut Laue-Langevin, 6 rue Jules Horowitz, 38042 Grenoble, France}

\author{A.~Schneidewind}
\affiliation{J\"ulich Centre for Neutron Science JCNS at Heinz Maier-Leibnitz Zentrum (MLZ), D-85748 Garching, Germany}

\author{R.~Georgii}
\affiliation{Heinz Maier-Leibnitz Zentrum (MLZ) and Physics Department E21, Technische Universit\"at M\"unchen, D-85748 Garching, Germany}

\author{O.~Zaharko}
\affiliation{Laboratory for Neutron Scattering and Imaging, Paul Scherrer Institute, CH--5232 Villigen, Switzerland}

\author{L.~Keller}
\affiliation{Laboratory for Neutron Scattering and Imaging, Paul Scherrer Institute, CH--5232 Villigen, Switzerland}

\author{Ch.~R\"uegg}
\affiliation{Neutrons and Muons Research Division, Paul Scherrer Institute, CH--5232 Villigen-PSI, Switzerland}
\affiliation{Department of Quantum Matter Physics, University of Geneva, 24, Quai Ernest Ansermet, CH--1211 Gen\`eve, Switzerland}

\author{C.~Pfleiderer}
\affiliation{Physics Department E51, Technische Universit\"at M\"unchen, D-85748 Garching, Germany}

\author{B.~Schmidt}
\affiliation{Max Planck Institute for the Chemical Physics of Solids, 01187 Dresden, Germany}

\author{N.~Shannon}
\affiliation{Okinawa Institute of Science and Technology Graduate University, Onna-son, Okinawa 904-0495, Japan}

\author{A.~Kriele}
\affiliation{German Engineering Materials Science Centre at MLZ, Helmholtz-Zentrum Geesthacht, Lichtenbergstr. 1, D-85748, Garching, Germany}

\author{A.~Senyshyn}
\affiliation{Heinz Maier-Leibnitz Zentrum (MLZ) and Physics Department E21, Technische Universit\"at M\"unchen, D-85748 Garching, Germany}

\author{A.~Smerald}
\affiliation{Max Planck Institut f\"ur Festk\"orperforschung, Quantum Materials Unit, D-70569, Stuttgart, Germany}
\date{\today}

\begin{abstract}

We report neutron scattering and AC magnetic susceptibility measurements of the 2D spin-1/2 frustrated magnet BaCdVO(PO$_{4}$)$_{2}$.
At temperatures well below $T_{\sf N}\approx 1K$, we show that only 34\% of the spin moment orders in an up-up-down-down strip structure. Dominant magnetic diffuse scattering and comparison to published $\mu$sr measurements indicates that the remaining 66\% is fluctuating.
This demonstrates the presence of strong frustration, associated with competing ferromagnetic and antiferromagnetic interactions, and points to a subtle ordering mechanism driven by magnon interactions.
On applying magnetic field, we find that at $T=0.1$ K the magnetic order vanishes at 3.8 T, whereas magnetic saturation is reached only above 4.5 T.
We argue that the putative high-field phase is a realization of the long-sought bond-spin-nematic state.

\end{abstract}


\maketitle

\begin{figure}[t]
\centering
\includegraphics[width=0.45\textwidth]{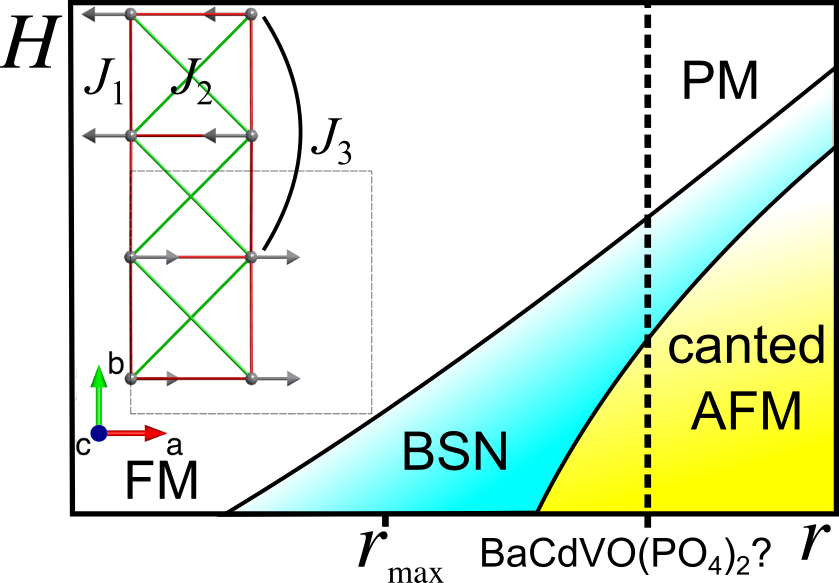}
\caption{\footnotesize{
Simplest generic phase diagram of 2D spin-1/2 frustrated magnets with competing ferromagnetic (FM) and antiferromagnetic (AFM) interactions (ratio $r$, maximal frustration at $r_{\sf max}$) and applied magnetic field ($H$),  showing paramagnetic (PM), bond spin nematic (BSN) and canted AFM phases \cite{shannon06,momoi06,ueda09,sindzingre09,sindzingre10,momoi12,ueda13,sizanov13,ueda15}.
The probable location of BaCdVO(PO$_{4}$)$_{2}$ is indicated.
(Inset) $H=0$, up-up-down-down spin ordering of BaCdVO(PO$_{4}$)$_{2}$ within a quasi-2D V layer, along with the simplest consistent magnetic exchange pattern, which has FM $J_1$ (red bonds), AFM $J_2$ (green bonds) and AFM $J_3$ (black bonds) Heisenberg interactions. 
}}
\label{fig:nematic_phasediag}
\end{figure}

%
In the search for new states of matter, the realization of a spin-nematic state --- a quantum version of a liquid crystal --- has proved an enduring but elusive goal \cite{andreev84}.
Of particular interest is the bond spin nematic (BSN), believed to exist in spin--1/2 materials 
with competing ferromagnetic (FM) and antiferromagnetic (AFM) interactions \cite{andreev84,shannon06}.
This state is remarkable in that it combines the long--range entanglement characteristic 
of a quantum spin liquid \cite{shindou09,shindou11,shindou13,benton2018}, with a nematic order 
that breaks spin--rotational symmetry, while preserving both translational-- and time--reversal 
symmetry \cite{andreev84,shannon06}.
As a consequence, 
the BSN state does not produce a static internal magnetic field, making it difficult to 
observe in experiment \cite{shindou13,smerald13,smerald15,smerald16}.
Nonetheless, there is now a well--established scenario for BSN order 
ocurring through the condensation of bound pairs of magnons in high magnetic field \cite{shannon06,momoi06,kecke07,hikihara08,sudan09,ueda09,sindzingre09,sindzingre10,zhitomirsky10,momoi12,ueda13,sizanov13,ueda15}, 
cf. Fig.~\ref{fig:nematic_phasediag}, and a number of promising 
candidate materials where this may occur \cite{svistov10,janson2016,nawa17,orlova17,grafe17}.


Of particular note is BaCdVO(PO$_{4}$)$_{2}$, one of a family of square--lattice, 
spin--1/2 vanadates \cite{melzi00,melzi01,kaul04,skoulatos07,nath08,skoulatos09}.
Early measurements of the heat capacity 
identified a phase transition with $T_{\sf N} \approx 1\ \text{K}$, for fields 
$H \leq 3.5\ \text{T}$, while the magnetization was found to saturate for 
a field $H \gtrsim 4\ \text{T}$ \cite{nath08}.
These results were interpreted in terms of a model with 
1$^{st}$--neighbor exchange $J_1 \approx -3.6\ \text{K}$ and 
2$^{nd}$--neighbor exchange \mbox{$J_2 \approx 3.2\ \text{K}$} \cite{nath08}, 
for which the low--field ordered state would be a canted AFM 
with propagation vector ${\bf q}_{\sf sq} = (1/2,0)$ \cite{shannon04,schmidt09,schmidt17a}.
Subsequent thermodynamic measurements have extended the 
magnetic phase diagram of BaCdVO(PO$_{4}$)$_{2}$, filling in the gaps
at high field, and identifying a new low--temperature phase  
bordering the saturated state \cite{povarov19}, 
precisely where one might expect a BSN to occur   \cite{shannon06,momoi06,ueda09,sindzingre09,sindzingre10,momoi12,ueda13,sizanov13,ueda15}.
This is an exciting development, since the range of fields involved, 
$4  \lesssim H   \lesssim 5\ \text{T}$ \cite{povarov19}, 
is much lower than in other candidate systems 
\cite{svistov10,janson2016,nawa17,orlova17}, making it 
accessible to a wider range of experimental techniques.
Despite this progress, the nature of the low--field phase in 
BaCdVO(PO$_{4}$)$_{2}$, the form of its magnetic interactions, 
and the possibility of a BSN in high field, all remain open questions.


In this Letter we use a combination of neutron scattering and AC susceptibility measurements 
to determine the magnetic behavior of BaCdVO(PO$_4$)$_2$ 
for fields ranging from $H=0$ to the saturated state for $H  \gtrsim 4.5\ \text{T}$.
We identify the ground state in zero field as an AFM with unusual 
``up--up--down--down'' order and small ordered moment, providing 
evidence both for strong frustration, and for interactions 
beyond a simple $J_1$--$J_2$ model.
We track the evolution of this state up to a field of 
$H_{\sf c}=3.8\ \text{T}$, where the associated Bragg peaks vanish.   
No new Bragg peaks appear at low temperature for $H > H_{\sf c}$, 
indicating that there is no new magnetic order.  
Nonetheless, the magnetization does not saturate, and AC susceptibility 
measurements reveal significant spin fluctuations, 
up to a field $H_{\sf sat} \gtrsim 4.5\ \text{T}$.
When combined with published thermodynamic measurements \cite{nath08,povarov19}, 
these results make a strong case for the existence of a BSN in BaCdVO(PO$_4$)$_2$,
in the field range $3.8\ \text{T} < H \lesssim 4.5\ \text{T}$.


{\it Experimental details:}
A powder sample of BaCdVO(PO$_{4}$)$_{2}$ was prepared as described in \cite{nath08}.
Since Cd is strongly absorbing neutrons, the isotope $^{114}$Cd was used, with $\sigma_{abs} = 0.34$ barns as opposed to 2520 barns for natural Cd.
%
%
%
%
%
The phase purity of the resulting powder was confirmed by x-ray diffraction using a D8 Advance Bruker AXS diffractometer.
%
%
The AC susceptometer consists of a primary excitation coil wound onto a high purity single crystalline sapphire tube.
A balanced pair of secondaries detect the sample signal. 
The sample was enclosed in a silver can and measurements were taken at a frequency of 120  Hz, down to $T=0.2$ K and up to $H=6$ T.
The energy integrated ($-12 \rm{meV}<E< 9 \rm{meV}$) magnetic signal was measured with polarized neutrons on D7, ILL, enabling the measurement of the magnetic structure factor $S(Q)$, where ${Q}$ is the momentum transfer.
Due to the low energy scale associated with magnetism in BaCdVO(PO$_{4}$)$_{2}$, the measurements capture the full spectral energy range \cite{skoulatos09,nath08}.
The IN5 instrument at ILL was also employed, with a dilution unit and $\lambda=8$ \AA, in order to reach lower $Q$ reflections.
Integration of the measured $S(Q,\omega)$, where $\omega$ is the energy transfer, was performed over an energy window of 0.01 meV (elastic line).
In-field experiments were also performed, with the DMC (PSI) and MIRA (FRM-II) \cite{georgii18} instruments, equipped with dilution units and vertical magnets up to 6 T.

\begin{figure}[t]
\centering
\includegraphics[width=0.45\textwidth]{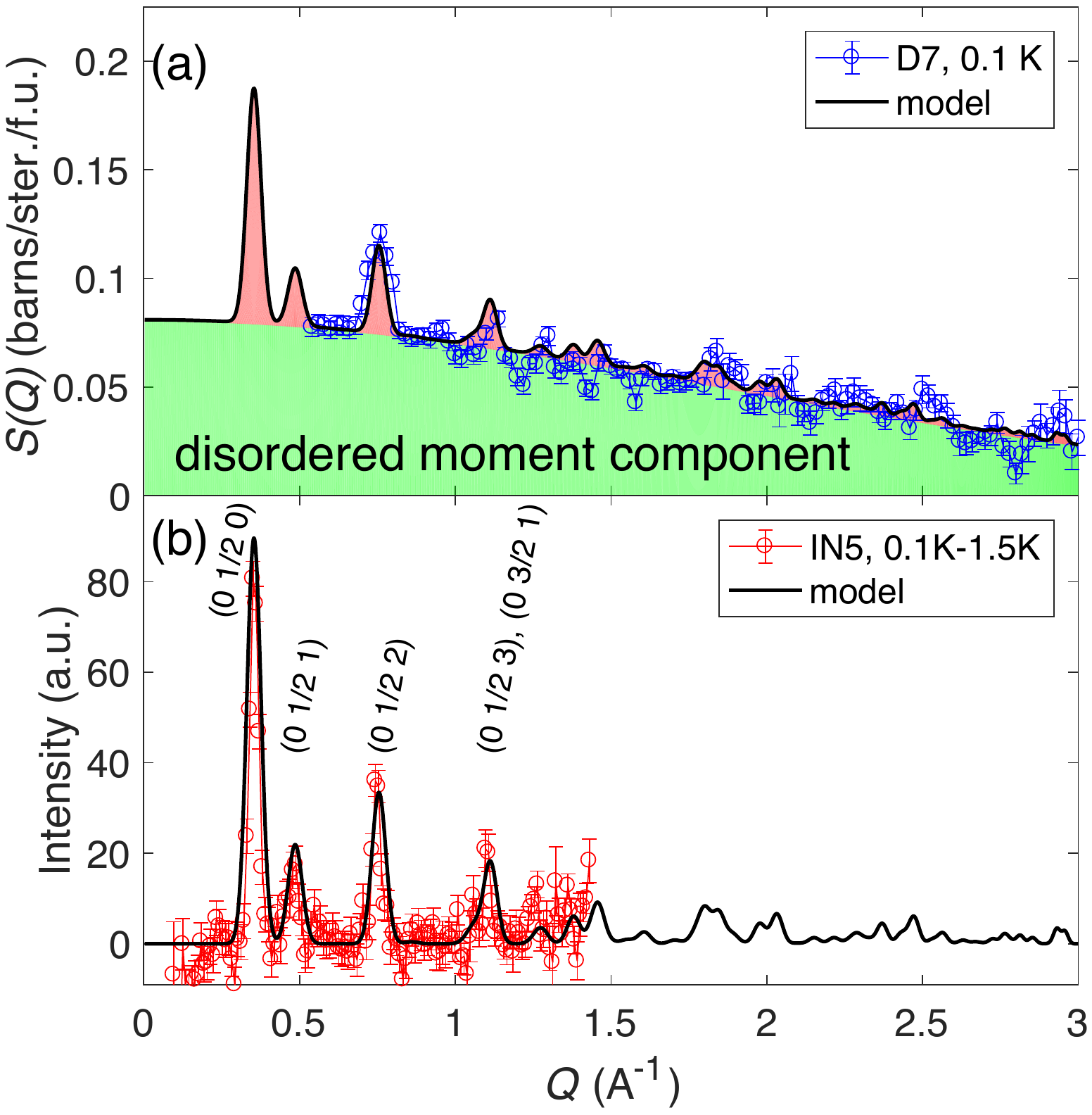}
\caption{\footnotesize{
Magnetic scattering in BaCdVO(PO$_{4}$)$_{2}$, as measured by neutrons at $H=0$ and at $T=0.1$ K.
Rietveld refinement is shown as a solid black line, with the most important peaks indexed. 
(a) D7 data showing purely magnetic scattering can be split into sharp magnetic peaks (red shading) and a diffuse background associated with the disordered component of the magnetic moment (green shading).
Analysis shows that only 34\% of the magnetic moment orders. 
(b) Elastic line of IN5 data allows access to lower $Q$ values.
Scattering at higher temperature has been subtracted, yielding the purely magnetic long range order.
}}
\label{fig2}
\end{figure}

{\it Magnetism at $H=0$:}
To probe the $H=0$ magnetism of BaCdVO(PO$_{4}$)$_{2}$, neutron scattering was performed at $T=0.1$ K and 1.5 K with two complementary instruments.
The $XYZ$ polarization analysis available on the D7 instrument allowed the purely magnetic part of $S(Q)$ to be measured, including the disordered component, with the results shown in Fig.~\ref{fig2}(a).
%
%
%
Since D7 was limited to $Q>0.5~\text\AA^{-1}$, IN5 measurements were also performed to access down to $Q>0.1~\text\AA^{-1}$.
The elastic structure factor $S(Q,\Delta E=0)$ was extracted and is shown in Fig.~\ref{fig2}(b), by subtracting the scattering at higher temperature (0.1 K -- 1.5 K).

%

Indexing of the magnetic Bragg peaks shows that the magnetic propagation vector is ${\bf Q} = (0,1/2,0)$ in the crystallographic unit cell of 8 vanadium ions ($2\times2\times2$).
Translating this propagation vector into the theoretical language of the square lattice of V ions, one finds ${\bf q}_{\sf sq} = (0,1/4)$, where ${\bf q}_{\sf sq}$ is defined relative to an idealized 1-site unit cell.
Thus our samples do not realize the columnar ${\bf q}_{\sf sq} = (0,1/2)$ magnetic order that was previously proposed from fitting high-temperature series expansions of a $J_1$-$J_2$ Heisenberg model to magnetic susceptibility measurements \cite{nath08} and assumed thereafter \cite{povarov19}.
In fact the propagation vector is incompatible with any of the phases of the $J_1$-$J_2$ Heisenberg model that have previously been assumed \cite{nath08,carretta09,roy11,tsirlin09a,tsirlin09b}.

The D7 instrument measures both the ordered and disordered components of the magnetic moment.
As shown in Fig.~\ref{fig2}a), only a small fraction of the magnetic moment gives rise to sharp magnetic peaks (red shaded area), while the majority gives rise to broad diffuse scattering.
Analysis of the ratio shows that 34\% of the total magnetic moment orders, a remarkably small fraction.
It is not {\it a priori} clear from neutron scattering whether the remaining 66\% is frozen or fluctuating.
However, previously published $\mu$sr measurements lend strong support to it being dynamic \cite{carretta09}.
Muon experiments at $T=0.35$ K show oscillations consistent with the presence of an ordered moment, but there is also a non-zero, temperature-independent longitudinal relaxation rate for $T<T_{\sf N}$, demonstrating persistent spin fluctuations at low temperatures \cite{carretta09,yaouanc15}.

The most obvious origin of a large fluctuating moment at $T\ll T_{\sf N}$ in a quasi-2D material is magnetic frustration.
This is confirmed by published susceptibility measurements, which show that both the Curie-Weiss temperature, $\theta_{\sf CW}$, and $T_{\sf N}$, are low compared to the magnetic energy scale of the system (temperature of the susceptibility maximum) \cite{nath08}.
This implies that the material has competing FM and AFM interactions, and our neutron measurements show that this frustration results in large quantum fluctuations at low-temperature.

\begin{figure}[t]
\centering
\includegraphics[width=\columnwidth]{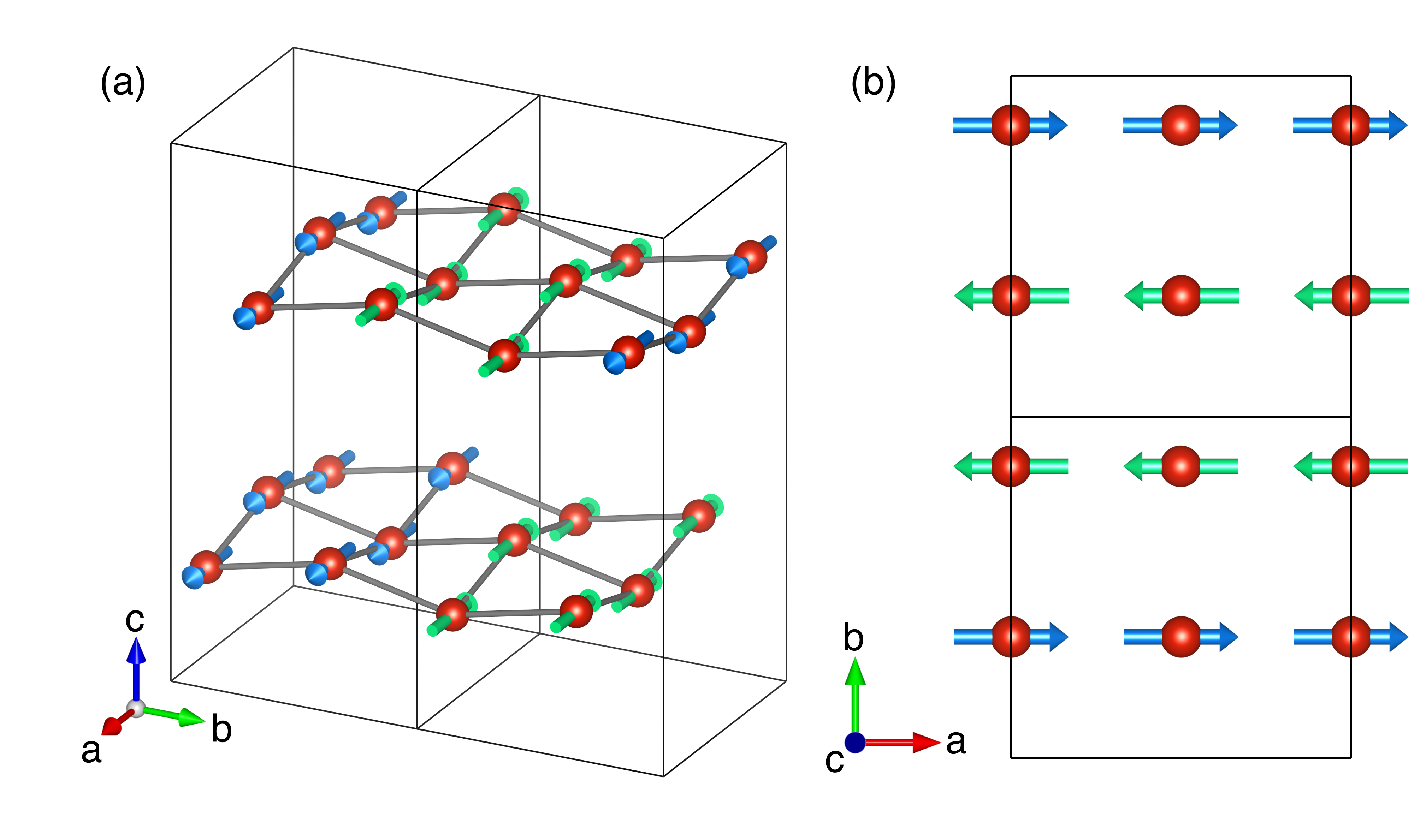}
\caption{\footnotesize{
Magnetic structure of BaCdVO(PO$_{4}$)$_{2}$ as determined by neutron diffraction.
The V$^{4+}$ ions form a distorted square lattice with ordered moments along the $a$ direction.
a) shows a general view and b) the $ab$ square--lattice plane.
The different spin colors differentiate between moment directions.
The structural unit cells are shown as solid-line boxes/squares. 
}}
\label{fig3}
\end{figure}

In order to determine the magnetic structure, the possible magnetic space groups and moment directions were determined using the MAXMAGN program on the Bilbao crystallographic server \cite{maxmagn,Aroyo:xo5013}.
Fitting the experimental data to the four resulting possibilities according to the standard Rietveld refinement procedure \cite{carvajal93} yields $P_bna2_1$ as the best solution.
There is a very good fit to the data for spins parallel to the $a$ axis, with the structure shown in Fig.~\ref{fig3}.
This is consistent with measurements of an anisotropic magnetic susceptibility \cite{povarov19}.
However, due to the relatively small number of peaks, we cannot definitively exclude the possibility that there is a small component of the moment parallel to the $b$ or $c$ axes.
The ordered moments form an up-up-down-down structure propagating along the $b$ axis, with an ordered moment $\boldsymbol{\mu}_a=0.34(3)\mu_{\sf B}$ (assuming pure spin-1/2).
In the $c$ direction the spin patterns in neighboring $ab$ planes are shifted with respect to one another by one lattice site, as a consequence of the magnetic symmetry.

The magnetic structure provides information about the magnetic Hamiltonian.
The up-up-down-down stripe arrangement in $ab$ planes shows that BaCdVO(PO$_{4}$)$_{2}$ is not described by the previously discussed $J_1$-$J_2$ Heisenberg model \cite{nath08,carretta09,roy11,tsirlin09a,tsirlin09b}.
In fact, while somewhat related structures have been proposed in various contexts, including $^3$He on an fcc lattice, where multiple ring exchange interactions are important \cite{roger83}, $J_1$-$J_2$ spin-1/2 chains, where it can be stabilized by Ising anisotropy \cite{igarashi89} and in the distorted spinel GeCu$_2$O$_4$ \cite{zou16}, the structure remains quite unusual.
The collinear (rather than spiral) nature of the spins points to the importance of quantum effects.

The simplest Hamiltonian consistent with the up-up-down-down structure and the isotropic nature of the exchange interactions in vanadates \cite{tsirlin09b}, is the square-lattice $J_1$-$J_2$-$J_3$ Heisenberg model with FM $J_1$ and AFM $J_2$ and $J_3$ (see Fig.~\ref{fig:nematic_phasediag}).
While a classical analysis of the model does not yield an up-up-down-down phase, exact diagonalization studies show that for spin-1/2 such a phase is stabilized by quantum fluctuations \cite{sindzingre09,sindzingre10}.
This occurs in a region of the phase diagram where $J_2+J_3$ is comparable to $J_1$, in accord with the finding of a small $\theta_{\sf CW}$.
The lack of classical stability of the ordering pattern means that it is not understandable within the non-interacting-magnon approximation of linear spin-wave theory, but is instead dependent on strong magnon-magnon interactions.
Understanding this subtle type of frustration-driven order is an interesting topic in itself.
However, the most interesting feature of the model is that on applying magnetic field, a spin-nematic phase is formed,  due to a strong binding energy between magnons \cite{sindzingre09}.


\begin{figure}[t]
\centering
\includegraphics[width=\columnwidth]{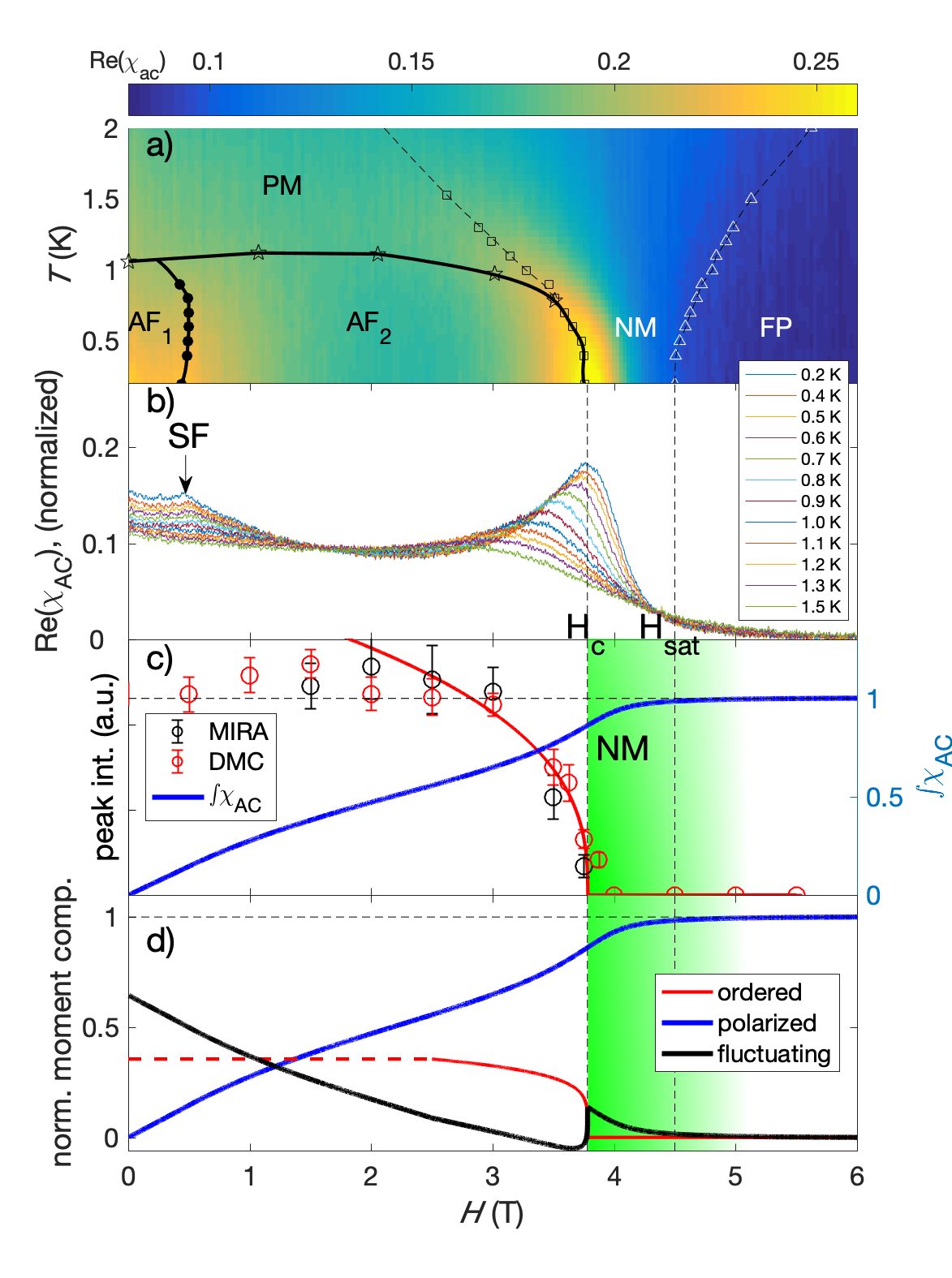}
\caption{\footnotesize{
$H-T$ phase diagram of BaCdVO(PO$_{4}$)$_{2}$ as determined by neutron scattering and AC magnetic susceptibility measurements.
a) At low temperature a powder-averaged spin-flop transition at $H\approx 0.44$ T (circles) separates low-field (AF$_1$) and intermediate field (AF$_2$) regions.
The field-induced transition out of AF$_2$ at low-$T$ takes place at $H_c\approx 3.76$ T (squares),  below the formation of a fully polarized state at $H_{\sf sat}\geq4.5$ T (lower bound, triangles).
In between lies a non-magnetic region (NM).
At $T\approx 1$ K the AF phase transitions into a paramagnet (PM, star symbols taken from \cite{nath08}).
b) Individual $H$ scans from which the phase diagram is deduced.
c) Field dependence of the ${\bf Q}=(0,1/2,0)$ magnetic reflection at $T=0.1$ K (circles) and integrated AC susceptibility (blue line).
$H_c=3.78(4)$ T is determined from a power law fit to the neutron data, while $H_{\sf sat}\geq 4.5$ T is the lower bound given by integrated AC susceptibility.
(d) Ordered (red), polarized (blue) and fluctuating (black) components of the magnetic moment.
}}
\label{fig4}
\end{figure}

{\it $H$-$T$ phase diagram:}
In order to understand the behavior of BaCdVO(PO$_{4}$)$_{2}$ in an applied magnetic field, field dependent neutron diffraction and AC magnetic susceptibility measurements were performed and used to map out the phase diagram shown in Fig.~\ref{fig4}(a).

Neutron diffraction was used to measure the field-dependent intensity of the ${\bf Q}=(0,1/2,0)$ reflection at $T=0.1$ K, with the results shown in Fig.~\ref{fig4}c.
At low field the measurements show an increasing peak intensity with increasing magnetic field (Fig.~\ref{fig4}c).
This is related to an observed re-entrant behavior in which $T_{\sf N}$ initially increases with field (Fig.~\ref{fig4}a).
Reentrance of this type is a sign of strong frustration-induced moment suppression \cite{schmidt17a,schmidt17b}.

The most important finding of the neutron measurements is that the ${\bf Q}=(0,1/2,0)$ magnetic reflection disappears at $H_{\sf c}=3.8$ T, and no new magnetic reflections appear for higher field.
Thus there is no magnetic long-range order for $H>H_{\sf c}$.

The neutron diffraction measurements can be compared to field-dependent AC susceptibility measurements, performed for temperatures 0.2 K$<T<$2.0 K, shown in Fig.~\ref{fig4}b.
The low temperature data shows an anomaly at $H= 0.44$ T, attributed to a spin-flop transition driven by a small exchange anisotropy.
%
The main peak at $H_{\sf c} \approx 3.8$ T, provides excellent corroboration of the phase transition observed in neutron scattering.
The close agreement can be excluded to be a fortuitious result of powder averaging, due to the small critical-field anisotropy observed in themodynamic measurements of single crystals \cite{povarov19}.
For $H\gtrsim H_{\sf c}$ the AC susceptibility signal remains strong, an inconsistent behavior for a fully polarized magnetic moment.
Instead the fading out of the AC susceptibility that signals the onset of full polarization has a conservative lower bound of  $H_{\sf sat} \geq 4.5$ T, and could be as high as 5 T.

Taking these results together allows the magnetic moment to be split into three components, as shown in Fig.~\ref{fig4}d.
The first is the ordered ${\bf Q}=(0,1/2,0)$ component, which is revealed by neutron scattering to account for 34\% of the total at $H=0$ and disappear at $H_{\sf c}=3.8$ T.
This has been fitted with a power law in the region $H>2.5$ T (Fig.~\ref{fig4}c).
The second is the polarized ${\bf Q}=0$ component, which can be determined from integration of the AC susceptibility.
%
The validity of this integration procedure is due to the vanishingly small imaginary component of the AC susceptibility (not shown).
The remainder of the moment fluctuates over all ${\bf Q}$ and $\omega$ and can be determined from what remains once the ordered and polarized moments have been taken into account.
It can be seen in Fig.~\ref{fig4}d that, starting at $H=0$, the fluctuating component steadily decreases with increasing $H$, before suddenly jumping at $H=H_{\sf c}$ and then gradually fading out with increasing field.
This is exactly what is expected for a spin-nematic state.


%
In conclusion, bringing together many different strands of evidence shows that BaCdVO(PO$_{4}$)$_{2}$ is a quasi-2D spin-1/2 magnet with competing ferromagnetic and antiferromagnetic interactions.
The resulting frustration drives strong quantum fluctuations around the low $T$, $H=0$ up-up-down-down ordered structure, leading to a remarkably small ordered moment.
Application of magnetic field fully suppresses the ordered moment at $H=3.8$ T.
No new magnetic ordering is detected for $H>H_{\sf sat}$, despite full polarization not occurring until $H_{\sf sat}>4.5$ T.
Instead, the non-polarized component of the magnetic moment fluctuates in ${\bf Q}$ and $\omega$.
Taken together, these results are strongly suggestive of a spin-nematic state in the region $H_{\sf c}<H<H_{\sf sat}$.
Definitive confirmation of the existence of a nematic state is likely to require probing the dynamic response of the material within the high-field phase \cite{smerald13,smerald15,smerald16}.

{\it Acknowledgments:}

We thank J.P. Goff, G. Benka, A. Bauer, A. Tsirlin, E.E. Kaul, C. Geibel, D. Sheptyakov, N. Qureshi, T. Fennell, S. Ward and S. T\'{o}th for useful discussions.
This research was supported by TRR80 project F2 of the German Research Foundation (DFG) and by NCCR MaNEP of the Swiss National Science Foundation (SNF).
Work was performed at the FRM II (DE), Institute Laue-Langevin (FR) and Swiss spallation source PSI (CH).


\bibliographystyle{apsrev4-1}
\bibliography{letter}

%

\end{document}